\documentclass[11pt]{article}
\usepackage{moriond,epsfig}
\usepackage{amsmath}
\usepackage{amsfonts}
\usepackage{amssymb}
\usepackage{graphics}
\usepackage{tabularx}

\newcommand{\circled}[1]{\raisebox{.5pt}{\textcircled{\raisebox{0pt} {{\scriptsize #1}}}}}
\newcommand{\MSbar}{\overline{\text{MS}}}

\def\be{\begin{equation}}
\def\ee{\end{equation}}
\def\bea{\begin{eqnarray}}
\def\eea{\end{eqnarray}}

\def\gsim{\:\raisebox{-0.5ex}{$\stackrel{\textstyle>}{\sim}$}\:}

\begin{document}
\begin{center}
\includegraphics[height=50mm]{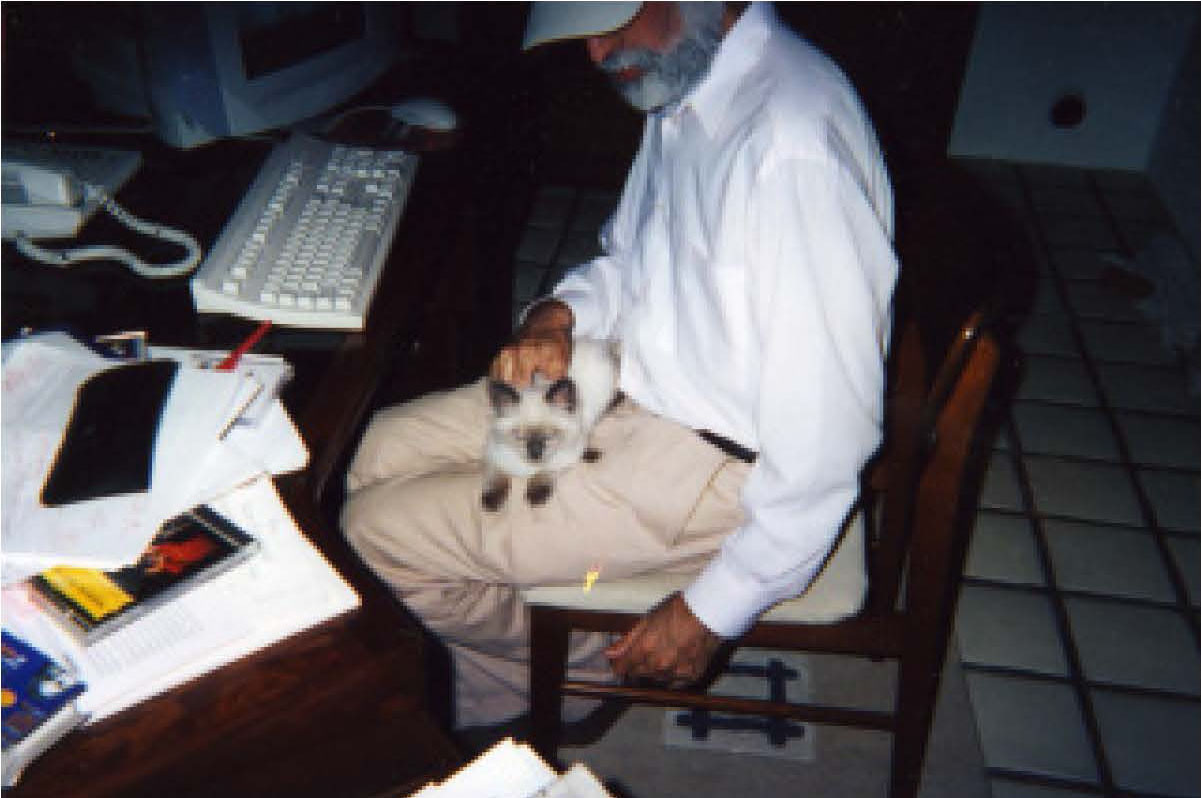}
\end{center}
\vspace*{1cm}
\title{Lattice understanding of the Delta I=1/2 rule \& some implications}\footnote{Invited
talk at the EW Moriond 2013}

\author{AMARJIT SONI}

\address{Department of Physics, Brookhaven National Laboratory,\\
 Upton, NY 11973, USA}

\maketitle
\abstracts{After decades of intensive efforts,
lattice methods finally revealed  one clear  source of the large  enhancement of the 
ratio $Re A_0/Re A_2$~\cite{RBC_UKQCD_PRL13}, which has been a puzzle in particle physics for about sixty years. Lattice studies of direct $K \to \pi \pi$ in the $I=2$ channel show that in fact this channel clearly suffers from a severe suppression due to a  significant cancellation
between the two  amplitudes for the original, charged current (tree) operator.
% [($\bar s_\alpha \gamma_mu (1 - \gamma_5) u_\alpha)(\bar u_\beta \gamma_mu(1-gamma_5)d_\beta$)],  
One of these amplitudes goes as N and the other one goes as $N^2$, where $N=3$ for QCD. For physical pion masses the cancellation between the two contributions towards $Re A_2$   is about 70\%. This appreciable cancellation suggests that expectations from large N for QCD may be amenable to receiving significant corrections.  The penguin operators  seem to make a small  contribution to $ReA_0$ at a scale $\gsim 1.5 GeV$.  Possible repercussions of the lattice observation for  other decays  are briefly discussed.}  
%Ref RBC_PRL
%LL operator

\section{Introduction}

%\subsection{Producing the Hard Copy}\label{subsec:prod}

Quantitative understanding of the long-standing $\Delta I=1/2$ puzzle and more importantly a reliable calculation
of the important direct CP-violation parameter in $K \to \pi \pi$, $\epsilon^\prime/\epsilon$  were in fact the primary
motivation for my entry into lattice methods for calculating weak matrix elements, about thirty years ago~\cite{BDHRS,BrGa,CMP,BDSPW,SS86}. Infact to tackle this difficult problem and bring it to our current level of understanding and progress has so far taken at least six 
Ph D theses~\cite{TD_th,CC_th,JL_th,SL_th,mL_th,QL_th}.
Indeed, at the time the experimental measurement
of $\epsilon'$ was  a huge challenge and  it took close to ~20 years to completely nail it down experimentally. 
For the lattice there were numerous obstacles that had to be overcome. First and foremost was lack of  chiral symmetry of Wilson fermions  entailing mixing with lower dimensional operators~\cite{Boc85,BDHS87}. While this severe difficulty thwarted early attempts
for all application to kaon physics (even for kaon-mixing parameter, $B_K$~\cite{BS88}), it motivated us to consider applications to
heavy-light physics as it was felt that therein chiral symmetry will be less of an issue~\cite{BDHS88,BES91,BLS93,BBS98}. 
Many of the important applications to
observables relevant to the Unitarity Triangle are in fact off springs of these efforts.

While the primary focus of this article is on developments exclusively  from the lattice perspective, we want to use the opportunity
to mention some prominent studies of $K \to \pi \pi$, the $\Delta~ I=1/2$ puzzle and $\epsilon'$  using continuum techniques
which offered interesting and very useful insights~\cite{cont}.

%\newpage

In 1996-97  the first simulations with domain wall quarks (DWQ) demonstrated the feasibility of using this 5-dimensional formulation~\cite{FS_94};  even with a modest extent of about 10 sites in the 5th dimension,  Domain Wall  Quarks (DWQ)  exhibited excellent chiral symmetry as the first application to kaon matrix element, in the quenched approximation,
showed~\cite{BS96,BS97}. With the formation of  RIKEN-BNL-Columbia (RBC) Collaboration around   ~1998 first large scale simulations, in the quenched 
approximations~\cite{RBC_cl},   with domain wall
quarks to $K \to \pi \pi$, $\Delta I=1/2$  and $\epsilon'$~  began. These continued to use chiral perturbation
theory (as was the case with the previous  attempts with Wilson fermions) to reduce the problem to
a calculation of $K \to \pi$ and $K \to vac$  following~\cite{BDSPW}.  The first results from this approach showed that
for $\epsilon'$, quenched approximation is highly pathological~\cite{RBC_eps01}. In particular, the QCD penguin operator 
$Q_6$ which is an (8,1) suffers from mixing with the (8,8) operators such as $Q_8$ emphasizing to us the need for full QCD in so far as the calculation of $\epsilon'$ is concerned~\cite{GP1,GP2,RBC_JL}.

It took several years to finish the first calculation of $K \to \pi \pi$ with DWQ  in full (2 + 1) flavor QCD again using ChPT only to discover that the kaon is simply too heavy for ChPT to be reliable~\cite{RBC_UKQCD_SU2}; the systematic errors for matrix elements of many of the key operators 
were O(50\%) or even more~\cite{SL_NHC,SL_th}. 

That brings us to the efforts of the past $\approx$ 6  years jointly by RBC and UKQCD collaborations to go instead  for {\it direct} calculations of $K \to \pi \pi$ using finite volume correlation functions as suggested by Lellouch-Luscher~\cite{LL00}. 
The results reported in this
talk~\cite{RBC_UKQCD_PRL13} are primarily using three different lattices (see  Tab.\ref{tab:params}) accumulated over the past several years. The $16^3$ and $32^3$ lattices only allow for threshold studies, whereas the $32^3$ lattice of volume ~  $(4.5 fm)^3$
is used to study  $K \to \pi \pi$ with physical kinematics. 
While all three lattices have been used already for the simpler I=2
final state, for the more challenging I=0 final state studies at physical kinematics on the $32^3$ lattice are still not
complete.  Fortunately, as will be explained, for the $\Delta I=1/2$ puzzle, it turns out that understanding the simpler $I=2$ channel proves to be crucial.

\subsection{The Puzzle}

Let's briefly recapitulate the so-called $\Delta I=1/2$ puzzle.
The issue  boils down to the huge (factor of about 450) disparity in the life-times of neutral ({\it i.e} $K_S$) 
and that of $K^\pm$. Thus, basically the spectator u-quark in $K^+$ is changing to d-quark in $K_S$ resulting in
this huge change in their life-times.   Their main decay mode is just to two pions. However, whereas $\pi^+ \pi^0$ (resulting from  the decays of $K^+$) is in
a pure $I=2$ final state, $\pi^+ \pi^-$ or $\pi^0 \pi^0$ are mixtures of $I=0$ and $I=2$; thus  
the ratio of the two relevant amplitudes $Re A_0/ReA_2 \approx 22$, for the  $I=0$  and $I = 2$ is  a lot bigger than
unity. Since  for the charged K
the change in isospin, $\Delta I=3/2$ whereas for the neutral K its either 1/2 or 3/2, it implies that the $\Delta I=1/2$
amplitude is significantly larger than the $\Delta I=3/2$ and this is the long-standing puzzle (see {\it e.g.}~\cite{DGH_book}).
While its long been speculated that QCD corrections may be responsible for this huge enhancement,
at this scale highly non-perurbative effects are anticipated; of course, over the years there have been numerous suggestions
~\cite{cont}, including new physics (see {\it e.g.}~\cite{AK94}) as the cause for this large enhancement.

\section{Weak Effective Hamiltonian and 4-quark operators}

Using the OPE apparatus, one arrives at the effective Hamiltonian for  $\Delta S=1$ weak decays~\cite{GMLR,AJBRMP,RBC_eps01},

\begin{equation}
 H^{\Delta S=1}=\frac{G_F}{\sqrt{2}}V_{ud}^*V_{us}
                      \sum_{i=1}^{10}[(z_i(\mu)+\tau y_i(\mu))] Q_i.
\label{eq: Eff_H}
\end{equation}

Here, $Q_i$ are the well-known 4-quark operators,

\begin{subequations}
  \allowdisplaybreaks[2]
\begin{align}
Q_1 &= (\bar{s}_\alpha d_\alpha)_{V-A}(\bar{u}_\beta  u_\beta )_{V-A}, \\
Q_2 &=  (\bar{s}_\alpha d_\beta)_{V-A}(\bar{u}_\beta  u_\alpha)_{V-A},\\
Q_3 &= (\bar{s}_\alpha d_\alpha)_{V-A}\sum_{q=u,d,s} (\bar{q}_\beta  q_\beta )_{V-A} ,         \\
Q_4 &=  (\bar{s}_\alpha d_\beta )_{V-A}\sum_{q=u,d,s} (\bar{q}_\beta  q_\alpha)_{V-A},\\
Q_5 &=  (\bar{s}_\alpha d_\alpha)_{V-A}\sum_{q=u,d,s} (\bar{q}_\beta  q_\beta )_{V+A}, \\
Q_6 &=  (\bar{s}_\alpha d_\beta )_{V-A}\sum_{q=u,d,s}(\bar{q}_\beta  q_\alpha)_{V+A} ,\\
Q_7 &= \frac{3}{2} (\bar{s}_\alpha d_\alpha)_{V-A}\sum_{q=u,d,s} e_q (\bar{q}_\beta  q_\beta )_{V+A},\\
Q_8 &= \frac{3}{2} (\bar{s}_\alpha d_\beta )_{V-A}\sum_{q=u,d,s} e_q (\bar{q}_\beta  q_\alpha)_{V+A},\\
Q_9 &= \frac{3}{2} (\bar{s}_\alpha d_\alpha)_{V-A}\sum_{q=u,d,s} e_q (\bar{q}_\beta  q_\beta )_{V-A},\\
Q_{10} &= \frac{3}{2} (\bar{s}_\alpha d_\beta )_{V-A}\sum_{q=u,d,s} e_q (\bar{q}_\beta  q_\alpha)_{V-A},
\end{align} 
\label{eq:ops}
\end{subequations}

\noindent where $\alpha$, $\beta$ are color indices and (V - A) means $\gamma_\mu ( 1 - \gamma_5)$. 

It is important to recognize that $Q_2$ is the original 4-quark (tree) operator of the basic charged current
weak decay, $[ \bar{s}_\alpha (\gamma_\mu (1 - \gamma_5) u_\alpha] [\bar{u}_\beta \gamma_\mu ( 1 - \gamma_5) d_\beta]$, conventionally written here in the Fierz transformed basis. When you swich on QCD,
$Q_2$ is not multiplicatively renormalizable and as was realized long ago~\cite{GL74,AM74}, it mixes with another tree operator
$Q_1$. On the other hand, $Q_3$ to $Q_6$ are the QCD penguin operators~\cite{SVZ75} and $Q_7$ to $Q_{10}$ are the electroweak (EW)
penguin operators~\cite{GW78,GW79}.

On the lattice, in the absence of exact chiral symmetry, each of these dim-6, 4-quark operator of the $\Delta S=1$ Hamiltonian can mix with lower
dimensional operators, {\it e.g} $\bar{s}d$, $\bar{s}\gamma_5d$, etc.  The effects of these mixings are purely
unphysical and need to be subtracted away. As you make the lattice spacing finer and 
move towards the continuum limit, these unphysial contributions tend to become huge and it  can become a very demanding and delicate subtraction, quite akin to 
fine tuning.  The chiral behavor of wilson fermions was so bad that original methods~\cite{BDSPW,KS96,Maiani87}, that were proposed to deal with such subtraction issues proved to be
quite inadequate. Because of the excellent chiral symmetry of DWQs, this became  by and large  a non-issue
provided the extent of the 5th dimension is not too small.

\begin{figure}[htb]
\centering
\includegraphics[width=8cm]{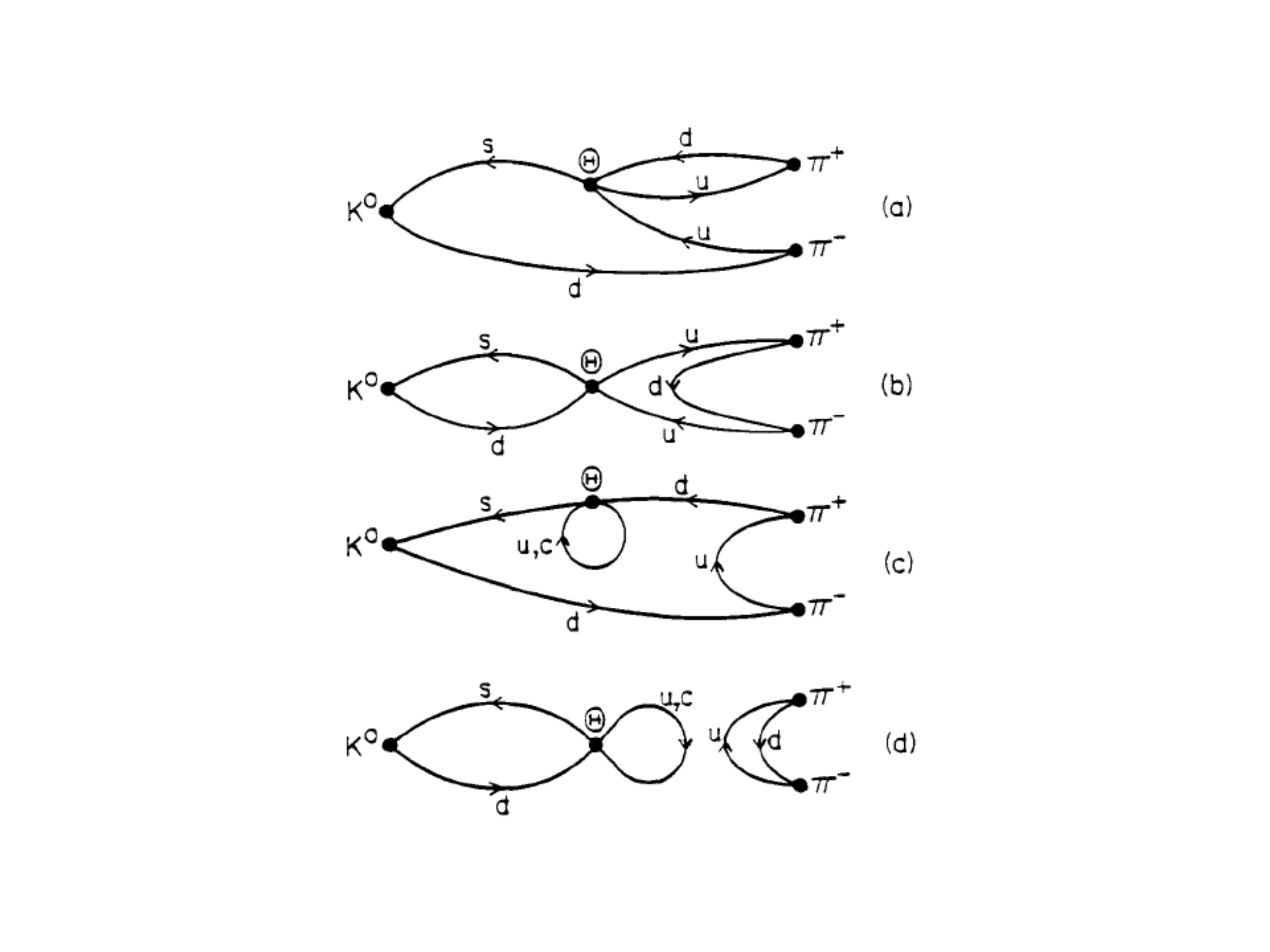}
\caption{Four general type of quark flow diagrams contribution to $K^0 \to \pi^+ \pi^-$; (a) corresponds
to spectator types in the continuum literature, (b) and (d) to annhilation and (c) to penguins; (d) though 
requires disconnected contributions which on the lattice are extremely demanding. Taken from~\protect\cite{BS88}}
\label{4_types} 
\end{figure}

Matrix element  $\langle\pi\pi|Q_i|K^0\rangle$ for each operator entail an evaluation of 48 different Wick contractions
which can be grouped into four different types~\cite{BS88,RBC_UKQCD_QL,QL_th}. Of these,  type-4 involve disconnected diagrams and are therefore, computationally the most demanding. Type-3 contain ``eye'' contractions~\cite{BDHRS}, type-2 correspond to
``figure-eight'' diagrams,
and type-1 correspond to original weak interaction tree graphs; see fig~\ref{4_types}. In particular, it is to be stressed that only type-1 contributes
to the $\Delta I=3/2$ transitions and the corresponding $I=2$ final state of the two pions whereas the
$\Delta I=1/2$ transtions for $I=0$ final state receive contributions from all four  types and consequently 
are much more  intricate and challenging  to tackle than the $\Delta I=3/2$  case.

\section{$\Delta I=3/2$}

As indicated already, ironically at the end of the day, it turned out that it is the simpler $\Delta I=3/2$,
$K \to \pi \pi$ that is very revealing in so far as the enhancement of the ratio is concerned. The 3/2  amplitude involves simply type-1 contractions~\cite{RBC_UKQCD_QL}. 
The Wick contractions for $\langle\pi\pi|Q_{1,2}|K^0\rangle$, for the dominant operators
$Q_2$ or $Q_1$, entail two contributions, one  goes as product of two traces in color space ($N \times N$) and 
the other is a 
single trace in color space (N), where for QCD, $N = 3$.

As is well known, continuum folklore says that $N^2$ term dominates and the two terms add~\cite{GL74,DGH_book_p,HG_book_p,CL_book_p}.
Our  data  using three different lattices~\cite{RBC_UKQCD_PRL13} collected over the past few  years allows us to study these contributions
as a function of the pion mass (with $m_K \approx 2 m_\pi$). In fact the relative sign between the terms is
negative and 
the cancellation between the two terms increases as the pion masses is lowered. Indeed at physical kinematics with $m_{\pi} = 142 MeV$
and $m_K = 520 MeV$,  the single trace contribution is around - 0.7 of the trace $\times$ trace term.
So, the observed amplitude is only around 2.7/12 $\approx$ 0.25 of naive expectations, assuming $N = 3$. In other words,
out of the observed enhancement in the ratio of the two amplitudes of a factor of around 22, as much
as a factor of 4 may simply be coming  from the fact that there is this cancellation making the 3/2
amplitude only about 0.25 of naive expectations.

\begin{figure}[htb]
\begin{minipage}{0.47\linewidth}
\centerline{\includegraphics[width=7cm]{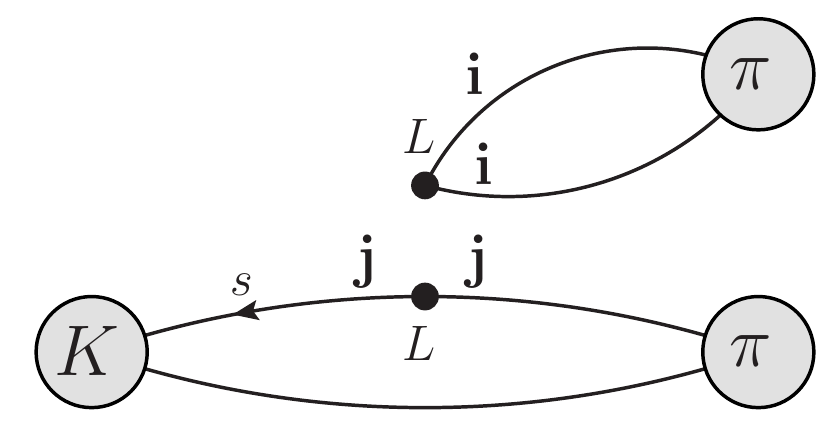}}
\end{minipage}
\hfill
\begin{minipage}{0.47\linewidth}
\centerline{\includegraphics[width=7cm]{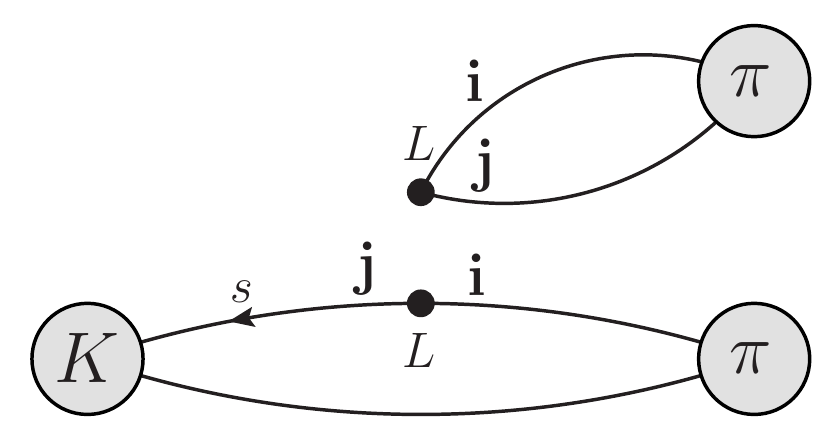}}
\end{minipage}
\caption{The two contractions contributing to Re$A_2$; $\bold i$ and $\bold j$ denote color indices. $s$ denotes the strange quark and $L$ that the currents are  left-handed; taken from~\protect\cite{RBC_UKQCD_PRL13}.} 
\label{c1c2}
\end{figure}

%\begin{subfigure}[h]{.238\textwidth}
%\includegraphics[width=\textwidth]{Fig1a_C1.pdf}
%\caption*{Contraction \protect\circled{1}.}
%\end{subfigure}
%\begin{subfigure}[h]{.238\textwidth}
%\includegraphics[width=\textwidth]{Fig1b_C2.pdf}
%\caption*{Contraction \protect\circled{2}.}
%\end{subfigure}
%\caption{The two contractions contributing to Re$A_2$. They are distinguished by the color summation ($\mathbf{i,j}$ %denote color). $s$ denotes the strange quark and $L$ that the currents are  left-handed. Taken from~\cite{RBC_UKQCD_PRL13.} 
%\label{c1c2}
%\end{figure}

Another notable feature of the $I = 2$ channel is that its amplitude, $ReA_2$, shows a significant dependence
on $m_{\pi}$. We attribute this largely to the cancellation mentioned above. From  Tab.\ref{tab:params}    we see that as the pion mass decreases from about 420 MeV to 140 MeV, $ReA_2$ decreases by about a factor of 3.5 and with physical $\pi$, K masses
it is in good agreement (within $\approx$ 15\%) with its measured value from experiments~\cite{RBC_UKQCD_PRD12}.

Moreover, recall that $Re A_2$ is closely related to $B_K$, the neutral Kaon mixing operator, as has been long known since
the famous work of ~\cite{DGH_BK}, who obtained $B_K^{LOChPT} \approx 0.3$ by exploiting its relationship with
the experimentally measured value of $ReA_2$ from the charged Kaon lifetime, assuming SU(3) and
lowest order chiral perturbation theory. Lattice studies for a long
time of course also have shown that $B_K$ changes from about 0.3 to 0.6 as you move  from the chiral limit to $m_K$~\cite{RBC_BKCL,LL11}.

\section{Implications for $ReA_0$ and the $\Delta I=1/2$ Rule}

What is even more striking is how this cancellation that is responsible for the suppression of 
$Re A_2$ actually also ends up enhancing $Re A_0$.
% The relationship between the leading operators and
%the two contractions, CN2 and CN1 ????????
First let's just look at the dominant operator, $Q_2$. Its contribution to
$Re A_2$ and to $Re A_0$ is as follows~\cite{RBC_UKQCD_QL}:

%\begin{equation}
%\label{eq:Q2_cont}
%\begin{subequations}
% \allowdisplaybreaks[2]
%\begin{align}
\begin{eqnarray}
Re A_{2,2} & = &i\sqrt{\frac{2}{3}}(ST +  TSQ),\\
Re A_{0,2} & = &i\sqrt{\frac{1}{3}}(-ST + 2 TSQ)
%\end{equation}
%\end{align}
\label{eq:Q2_cont}
\end{eqnarray}
%\end{subequations}

\noindent where $A_{(i,j)}$ notation means $ i=0$ or $2$,  (depending on the isospin of the  pion final state) and $ j=1,2$ and $j=2$, for example, means $Q_2$ and ST  means single trace over color indices and TSQ means trace $\times$ trace.   Thus, recalling
that at physical kinematics, $ST/TSQ \approx - 0.7$, the ratio, $Re A_0 /Re A_2 \approx 6.4$. So far
we only looked at the contribution of the dominant tree operator $Q_2$. Let us next, also
retain the next most important operator, which happens to be the tree operator, $Q_1$. One finds,

%\begin{equation}
%\label{eq:Q1_cont}
%\begin{subequations}
% \allowdisplaybreaks[2]
%\begin{align}
\begin{eqnarray}
Re A_{2,1} & = &i\sqrt{\frac{2}{3}}(ST + TSQ),\\
Re A_{0,1} & = &i\sqrt{\frac{1}{3}}(2 ST -  TSQ).
%\end{equation}
%\end{align} 
\label{eq:Q1_cont}
%\end{subequations}
\end{eqnarray}

\noindent Thus, incorporating the Wilson coefficients ($Z_j$, with $j = 1,2$)  for these two operators~\cite{RBC_UKQCD_QL},

\noindent $Z_1 = -0.30$ and $Z_2 = 1.14$,  one gets,

\begin{equation}\label{eq:ReA1A2}
Re A_i  = Z_j A_{i,j} \\
\end{equation}

\noindent for $i=0, 2$ corresponding to  $I = 0, 2$ for the two final states. Then given ST $\approx$ -0.7 $\times$ TSQ,
we get $ReA_0/ReA_2 \approx 10.8$; thus  accounting for almost half  of the experimental number
$\approx$ 22.5. 

Note  also that the cancellation between the single color trace and trace square term 
ends up causing not only a further suppression of ReA2 because of the fact that the sign of Wilson coefficient 
$Z_1$ is negative to that of $Z_2$, but in addition as an interesting  coincidence, it ends up enhancing $ReA_0$.
This is easily understood from the above simple eqns \ref{eq:Q2_cont}, \ref{eq:Q1_cont} as the relative signs
between single trace and the squared trace switch from  $Re A_2$  to $Re A_0$. 

%From the above equations  one can see that the cancellation among single trace and trace square terms
%ends up resulting in yet another partial cancellation for $Re A_2$ as the fact that the Wilson coefficients for
%the subdominant operator ($Q_1$) being negative renders now the contribution to $ReA_2$ 
%from $Q2$ to cancel against that of $Q_1$. In contrast the contributions of $Q_1$ and $Q_2$ add for
%the case of $Re A_0$. 

While our calculation of $ReA_0$ at physical kinematics is not yet complete, there are several interesting features
of the existing calculations summarised in  Tab.\,\ref{tab:params}  that are noteworthy.  One item to note is the ratio
$ReA_0/ReA_2$ resulting from our two completed calculations on the $16^3$ and $24^3$ lattices. It is 9
and 12 respectively. These numbers are for amplitudes calculated at threshold. As commented before
the corresponding $ReA_2$ on these lattices are factors of $\approx$ 3.5 and $\approx$ 2 times
the value of $Re A_2$ at physical kinematics. This is mostly the result of significant mass dependence 
of $Re A_2$. In contrast, our numbers for $Re A_0$ show  milder mass dependence and infact
the value we obtain on our $24^3$ lattice (whose volume is about three times bigger compared
to the smaller lattice) at threshold  is quite consistent with experiment;  whether this feature will remain true
at physical kinematics or not remains to be seen.

\begin{table*}[h!]
\caption{Reproduced from ~\protect\cite{RBC_UKQCD_PRL13}. Summary of simulation parameters and results obtained on three domain wall fermion ensembles. The errors with the Iwasaki action are statistical only, the second error for Re$A_2$ at physical kinematics from the IDSDR simulation is systematic and is dominated by an estimated 15\%
discretization uncertainty as explained in ~\protect\cite{RBC_UKQCD_PRD12}.}
\label{tab:params} 
\vspace{0.4cm}
\begin{center}
\footnotesize
\begin{tabular}{c | c c c c c c | l}
&$~a^{-1}$&$m_\pi$ & $m_K$ & Re$A_2$ & Re$A_0$ &$\frac{\mathrm{Re}A_0}{\mathrm{Re}A_2}$ & notes \\
&$[\text{GeV}]$&$[\text{MeV}]$ & $[\text{MeV}]$ & $[10^{\textrm{-}8}\, \text{GeV}]$ & $[10^{\textrm{-}8}\, \text{GeV}]$ &  &  \\
\hline
$16^3$ {\bf Iwasaki} &1.73(3)& 422(7) & 878(15) &4.911(31) &45(10)&9.1(2.1) &\,threshold calculation\\
$24^3$ {\bf Iwasaki} & 1.73(3)& 329(6) & 662(11) &2.668(14) &32.1(4.6)&12.0(1.7) &\,threshold calculation \\
${\bf IDSDR}$ & 1.36(1)& 142.9(1.1) & 511.3(3.9) &1.38(5)(26)&-& - &\,physical kinematics \\
${\bf Experiment}$ &--& 135\,-\,140&494\,-\,498&1.479(4)&33.2(2)&22.45(6)& 
\end{tabular}
\end{center}
\end{table*}

In passing let us note that from SU(2) ChPT description of $K \to \pi \pi$ one also finds
a significant dependence on pion mass of $ReA_2$ than of $ReA_0$ ~\cite{BC09} in qualitative
agreement with the lattice observations. 

\begin{figure}[h!]
\centering
\includegraphics[width=9cm]{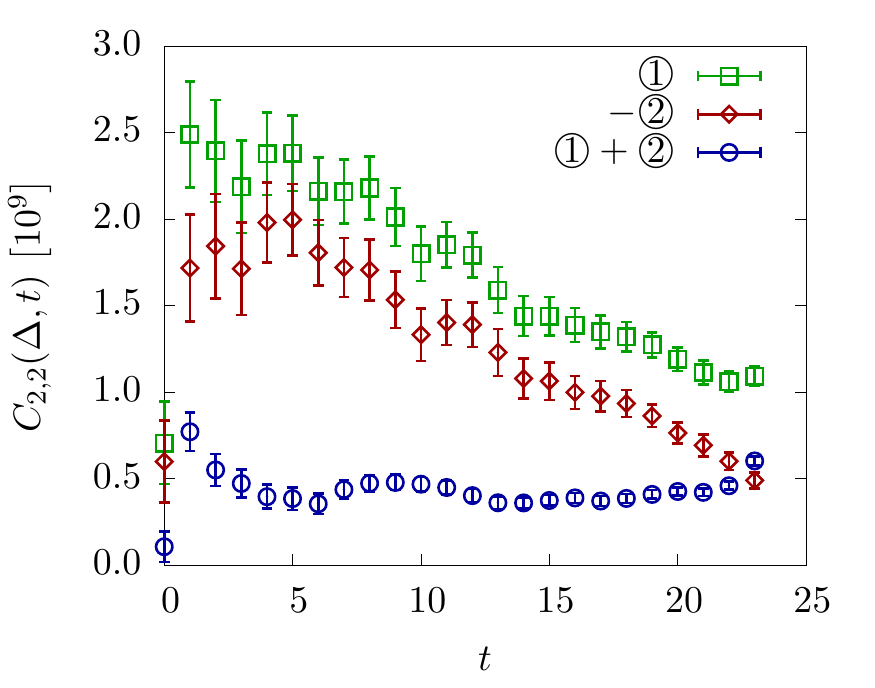}
%\includegraphics[width=.45\textwidth]{Fig2_c1c2data_DSDR_p2_dt24,heiht=3in}
%\pdffig{figure=Fig2_c1c2data_DSDR_p2_dt24,heiht=3in}
\caption{Contractions $\protect\circled{1}$ (that goes as trace $\times$ trace in color space and is also
call TSQ in here), -$\protect\circled{2}$ (that goes as single trace in color space and is also
called ST in here)  and $\protect\circled{1}+\protect\circled{2}$ as functions of $t$ from the simulation at physical kinematics. Taken from~\protect\cite{RBC_UKQCD_PRL13}. \label{c1c2physical}}
\end{figure}

\begin{figure}[h!]
\centering
\includegraphics[width=9cm]{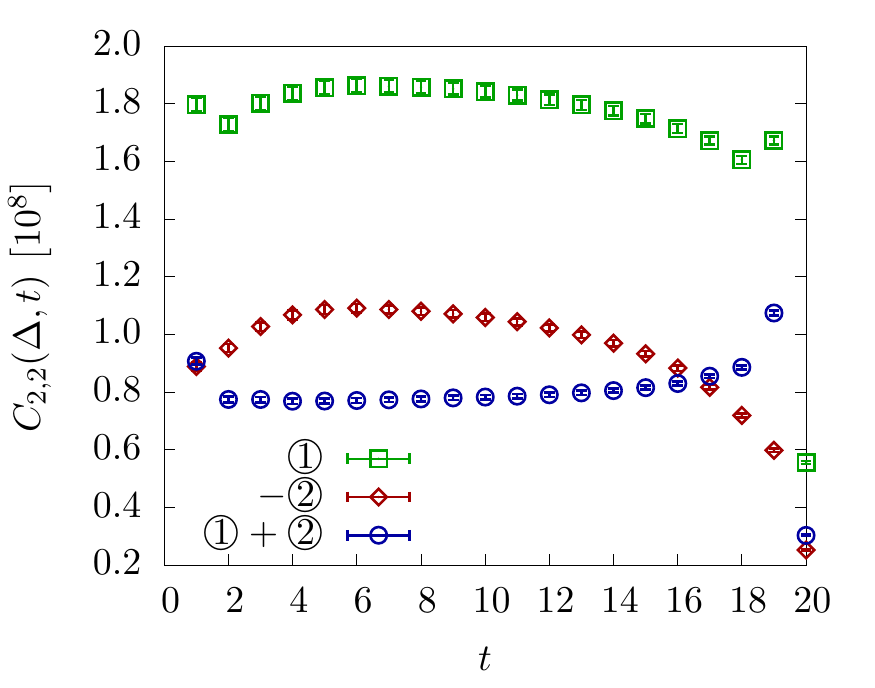}
%\includegraphics[width=.45\textwidth]{Fig3_c1c2data_24cube_dt20,height=3in}
%\includegraphics
%\pdffig{figure=Fig3_c1c2data_24cube_dt20,height=3in}
\caption{Contractions $\protect\circled{1}$, -$\protect\circled{2}$ and $\protect\circled{1}+\protect\circled{2}$ as functions of $t$ from the simulation at threshold with $m_\pi\simeq$ 330\,MeV; see also fig~\ref{c1c2physical}. Taken from~\protect\cite{RBC_UKQCD_PRL13}. \label{c1c2330}}
\end{figure}

\subsection{The role of penguins in the $\Delta I=1/2$ Puzzle}

%In the discussion above we have only mentioned the tree operators $Q_1$ and $Q_2$.
%Since these tree operators alone can give $R_{tree} \approx 10.8$ to 
%the observed ratio of $\approx$ 22, it should be clear that penguins
%can at best correct this by about a factor of two. 

From  Tab.\,\ref{tab:a0breakdown}  we see that at a scale of $\approx$ 2.15  GeV, the tree operators 
$Q_2$ and $Q_1$ account for almost 97\% of $ReA_0$ so the contribution of the remaining 
operators,  in particular the QCD penguins, is only a few \%
and  the EW penguins around 0.1\%. In fact roughly similar conclusions were arrived previosuly
when we used the chiral perturbation approach  both in the quenched approximation~\cite{RBC_eps01} 
as well as in dynamical 2+1 flavor QCD~\cite{SL_NHC,SL_th}. 

We stress again that these calculations~ Tab.\,\ref{tab:a0breakdown}
for $Re A_0$ are not at physical kinematics so the relative importance of the penguin to tree contributions
may well change to some degree; however, the fact remains that the cancellation and suppression of 
$ReA_2$ and enhancement of $ReA_0$,  in the tree contributions, which are the new aspects being reported
here, imply a diminished role for the penguin contributions at least at a renormalization point
around 2 GeV.
\begin{table}[h!]
\caption{\label{tab:a0breakdown} 
Contributions from each operator to Re$A_0$ for $m_K=662$\,MeV and $m_\pi=329$\,MeV. The second column contains the contributions from the 7 linearly independent lattice operators with $1/a=1.73(3)$\,GeV and the third column those in the 10-operator basis in the $\MSbar$-$\mathrm{NDR}$ scheme at $\mu=2.15$\,GeV. Numbers in parentheses represent the statistical errors.Taken from~\protect\cite{RBC_UKQCD_PRL13}}
\begin{center} 
\vspace{0.2cm}
\begin{tabular}{c|c|c}
i & $Q_i^{\text{lat}}\;$[GeV]&$Q_i^{\MSbar \text{-NDR}} \;$[GeV]\\ \hline
1&\,\phantom{-}8.1(4.6) $10^{-8}\,\,$&\phantom{-}6.6(3.1) $10^{-8}\,\,$\\ 
2&\,\phantom{-}2.5(0.6) $10^{-7}\,\,$&\phantom{-}2.6(0.5) $10^{-7}\,\,$\\ 
3&\,-0.6(1.0) $10^{-8}\,\,$&\phantom{-}5.4(6.7) $10^{-10}$\\ 
4&--&\phantom{-}2.3(2.1)~$10^{-9}\,\,$\\ 
5&\,-1.2(0.5) $10^{-9}\,\,$&\phantom{-}4.0(2.6) $10^{-10}$\\ 
6&\,\phantom{-}4.7(1.7) $10^{-9}\,\,$&-7.0(2.4) $10^{-9}\,\,$\\ 
7&\,\phantom{-}1.5(0.1) $10^{-10}$&\phantom{-}6.3(0.5) $10^{-11}$\\ 
8& -4.7(0.2) $10^{-10}$&-3.9(0.1) $10^{-10}$\\ 
9&--&\phantom{-}2.0(0.6) $10^{-14}$\\ 
10& --& \phantom{-}1.6(0.5) $10^{-11}$\\ \hline
Re$A_0$ &\,\phantom{-}3.2(0.5) $10^{-7}\,\,$ & \phantom{-}3.2(0.5) $10^{-7}\,\,$
\end{tabular}
\end{center}
\end{table}

\subsection{The role of disconnected diagrams for $Re A_0$}

Our calculation of $A_0$ being discussed here is not yet at physical kinematics. It is actually at threshold
and perhaps more importanty the (valence) 
pion masses $\approx$ are relatively heavy. As Tab.\,\ref{tab:params} shows we completed the threshold calculation 
of $Re A_0$ with  two different lattices ($16^3$ and $24^3$) with pion mass around 420 MeV and 330 MeV
attaining statistical accuracies around 25\% and 15\% respecitively. These calculations include the
contribution from disonnected diagrams as well. 
Within the stated accuracy, we do not seem to see any discernible contribution from
the disconnected diagrams in so far as $Re A_0$ is concerned. Again we emphasize that this is with pion mass around 330 MeV and
not with physical pion masses.

Given that the dominant contribution to $Re A_0$ seems to come from tree operators, 
which do not receive contribution from any disconnected diagrams, it is understandable that the
disconnected diagrams contribution to $Re A_0$ is most likely rather small.

\subsection{The role of disconnected diagrams for  $Im A_0$}

Tree level operators cannot contribute to $Im A_0$. Only eye-contractions and disconnected 
diagrams make contributions to $Im A_0$; thus one expects an enhanced role for disconnected graphs in $Im A_0$.
This is why our calculation of $Im A_0$, even with $m_\pi \approx 330 MeV$ has statistical errors of around 50\%.

\subsection{Status of $\epsilon^{\prime}$}

As is well known contributions to $\epsilon'$ can be divided into two categories: QCD penguins and EW penguins~\cite{GMLR,BJ04}
originating respectively from $Q_3$,$Q_4$,$Q_5$ and $Q_6$ and $Q_7$, $Q_8$, $Q_9$ and $Q_{10}$.
Amongst these $Q_6$ and $Q_8$ are the dominant players.

RBC and UKQCD have already finished their computation of $Im A_2$ as indicated in Tab.\,\ref{tab:params}  at physical
kinematics~\cite{RBC_UKQCD_PRD12} with an estimated statistical error of $\approx$ 20\% and  roughly similar error
for  systematics. This means the EWP contributions to $\epsilon'$ has already been completed; indeed improved calculations of $Im A_2$ are well underway
and are expected rather soon with appreciable reduction in errors.

From a purely personal perspective, $\epsilon'$  has always been the main focus of the $K \to \pi \pi$ effort from
the very beginning. The calculation of $Im A_0$ relevant to $\epsilon'$ is even more challenging than that of
$Re A_0$ relevant for the $\Delta I=1/2$ rule. This is because $Im A_0$ does not receive any contribution
from the tree operators. This is understandable as in the SM all three generations have to participate
to make a non-vanishing contribution to any CP violation phenomena. Thus penguin graphs
and consequently eye contractions become essential on the lattice.
While that renders the calculation quite challenging, perhaps another order of magnitude in the
complexity is added by the fact that the $I=0$ channel receives contributions from disconnected
diagrams. The error on our $\epsilon'$ calculaion is around 100\% at present.

\subsection{Possible implications for other weak decays}
Our lattice studies of {\it direct} $K \to \pi \pi$ seem to show that for QCD {\it i.e}, N=3, large N approximation
is amenable to rather large corrections. Since its use, as well as that of the closely related notion of factorization, is so pervasive in weak decays, perhaps, D and B decays ought to be re-examined in light of these lattice findings. Moreover, because the cancellation discussed above results in a significant fraction
of the enhancement of $Re A_0/Re A_2$ and also because the penguin contribution to $Re A_0$ seems to be so small,
it tells us that the  penguin contribution in D-decays (in the I=0 channel) is bound to be quite small as

\begin{equation}
\frac{P_D}{T_D} \approx  \delta_{Uspin} \times \frac{P_K}{T_K}
\label{eq:charm}
\end{equation}

\noindent  where, subscript D or K means exclusive $D \to PP$ or $K \to PP$, respectively, with P=pseudoscalar and 
$\delta_{Uspin}$ is indicative of Uspin violation reflecting the cancellation between the s
and d virtual quarks in the penguin. 

%This should be contrasted with expectations along the lines of ~\cite{Gr_Go}

\section{Summary \& Outlook for the near future}

Summarizing, lattice studies of {\it direct} $K \to \pi \pi$ show that in the simpler $I=2$ channel,
at physical kinematics,
the contributing amplitude from the original, tree, 4-quark $\Delta S=1$ weak operators that goes as 
$N^2$ cancels significantly with the one that goes as N, causing an appreciable suppression of the
$\Delta I=3/2$ transition~\cite{BBG_PR}. This seems to lead to a considerable  fraction of the enhancement in the 
ratio of $Re A_0/Re A_2$. These results suggest that expectations from large N may receive 
large corrections for QCD in weak decays.

Understanding $K \to \pi \pi$ decays and calculation of $\epsilon'$ remains a very important goal of the RBC
and UKQCD collaborations. I am hopeful that the first calculation of $Re A_0$ and $\epsilon'$ in full QCD
with physical kinematics would  be completed in about two years.

\section*{Acknowledgements}
I want to thank my colleagues from RBC and UKQCD collaborations and especially Norman Christ, 
Christoph Lehner, Qi Liu and Chris Sachrajda
for numerous valuable discussions. I have also benefitted from many conversations with Andrzej Buras
and Enrico Lunghi.
Also I must  thank the organizers of the EW Moriond 2013, and in
particular Jean-Marie Fr\`ere for inviting me.
This work is supported in part by the US DOE contract No.
DE-AC02-98CH10886.


\begin{thebibliography}{99}
\bibitem{BBG_PR}
	
	Remarkably, in ~\cite{cont}, the papers of Bardeen, Buras and Gerard and of Pich and Rafael, 
	written long ago, seem also
	to suggest some suppression of $Re A_2$. However, these papers make appreciable use of large N.
	On the other hand, as emphasized in here in our lattice calculation, the suppression of $ReA_2$, clearly  results  from  substantial cancellation between the leading and the nonleading terms in the $1/N$ expansion implying significant corrections to large N expectations. Therefore, the relation of these papers to our lattice calculations is unclear, at least to this author.
\end{thebibliography}
\end{document}